\documentclass[a4paper]{article}

\usepackage{INTERSPEECH2022}
\usepackage{multirow}
\usepackage{array}
\usepackage{amsmath,graphicx}
\title{Partial Coupling of Optimal Transport for Spoken Language Identification}
\name{Xugang Lu$^{1*}$, Peng Shen$^{1}$, Yu Tsao$^{2}$, Hisashi Kawai$^1$}
\address{
  $^1$National Institute of Information and Communications
Technology, Japan.\\
  $^2$Research Center for Information Technology Innovation, Academic
Sinica, Taiwan}
\email{xugang.lu@nict.go.jp}
\begin{document}
\maketitle
\begin{abstract}
In order to reduce domain discrepancy to improve the performance of cross-domain spoken language identification (SLID) system, as an unsupervised domain adaptation (UDA) method, we have proposed a joint distribution alignment (JDA) model based on optimal transport (OT). A discrepancy measurement based on OT was adopted for JDA between training and test data sets. In our previous study, it was supposed that the training and test sets share the same label space. However, in real applications, the label space of the test set is only a subset of that of the training set. Fully matching training and test domains for distribution alignment may introduce negative domain transfer. In this paper, we propose an JDA model based on partial optimal transport (POT), i.e., only partial couplings of OT are allowed during JDA. Moreover, since the label of test data is unknown, in the POT, a soft weighting on the coupling based on transport cost is adaptively set during domain alignment. Experiments were carried out on a cross-domain SLID task to evaluate the proposed UDA. Results showed that our proposed UDA significantly improved the performance due to the consideration of the partial couplings in OT.
\end{abstract}

\section{Introduction}
 Due to the success of deep learning (DL) technique in image and speech recognition, DL based algorithms have been proposed for spoken language identification (SLID) with significant performance improvement. The success of DL in SLID is mainly due to its effective language feature representation learning with large quantity of training data samples. For example, in current state of the art system for SLID, the feature representation based on X-vector which was inspired from speaker embedding \cite{Snyder2018}, could be extracted from a well trained deep neural network for language recognition with thousands of hours of speech. Based on the robust language feature representation, the SLID could be designed with a classifier model, either a conventional classifier model (e.g., Gaussian mixture model, logistic regression model) or another neural network based classifier model. There is also a unified modeling strategy, i.e., feature extraction and classifier are optimized simultaneously in an end-to-end neural network model for SLID \cite{RichardsonIEEE,Moreno2016,Moreno2014,Diez2015,Fernando2017,Geng2016}.

 In most studies, there is a simple assumption that the training and test data are independent and identically distributed (i.e., i.i.d.). However, it is usually the case that the recording environments of test data are different from those of training data. These difference may cause distribution mismatch between training and test conditions which will degrade the performance drastically. Although data augmentation in feature extraction and classification models with large range of noisy conditions could relieve the domain mismatch problem in some degree, it is impossible to cover all unknown test domain conditions. Domain adaptation is an indispensable solution for this domain mismatch problem. In addition, in real applications, the label information of a test data set is often unknown, therefore, unsupervised domain adaptation (UDA) is preferable. There are many UDA methods to deal with cross-domain mismatch problem, for example, feature-based correlation alignment (CORAL) \cite{LeeICASSP2019}, and feature-distribution adaptation \cite{Bousquet2019}. One of the most popular adaptation framework is based on maximum mean discrepancy (MMD) to reduce the domain feature distribution difference \cite{MMD}. With neural network learning algorithms, domain adversarial learning has been proposed to reduce feature domain discrepancy \cite{Ganin2015}. The purpose of these learning algorithms is to align the probability distributions between source and target domains.

 Concerning the probability distribution alignment or transform, optimal transport (OT) provides natural mathematic formulations, and has been intensively applied in machine learning field \cite{CourtyIEEE2017}. The initial motivation for OT in machine learning is to find an optimal transport plan to convert one probability distribution shape to another shape with the least effort \cite{Peyre2018}. By finding the optimal transport, it naturally defines a distance measure between different probability distributions. Based on this property, the OT is a promising tool for domain adaptation in image processing, classification, and segmentation \cite{CourtyIEEE2017,CourtyNIPS2017, DamodaranECCV2018}, as well as domain adaptation in speech enhancement \cite{LinNIPS2021}. Inspired by the OT based unsupervised adaptation \cite{CourtyIEEE2017,CourtyNIPS2017, DamodaranECCV2018}, we previously proposed an unsupervised neural adaptation framework for cross-domain SLID task \cite{ICASSPLu2021}. Based on the adaptation model, significant improvements were obtained on a cross-domain SLID task.

 In our previous work, we assumed that the training and test data domains share the same label space. However, in our SLID task, the labeling space of the test set is only a subset of that of the training set, i.e., training data set has 10 label categories, but the test data set only includes 6 label categories which is a subset of the 10 label categories. Reducing the domain discrepancy by fully matching their distributions may result in negative effect. In another word, in doing OT, training data samples which categories are not included in the test data set should not be coupled for distribution transport. Based on this consideration, in this paper, we further propose a joint distribution alignment (JDA) model based on partial optimal transport (POT) (hereafter JDA-POT) for SLID. Our main contributions are: (1) we proposed an JDA model based on partial coupling of OT for SLID via a neural network modeling framework. In distribution alignment based on the OT, only partial couplings between training and test are allowed, i.e., by setting a threshold of transport cost, training samples with transport cost larger than the threshold will not join in the distribution alignment.  (2) In order to reduce the risk of removing possible matched training samples, a soft weighting on the coupling is adaptively set during distribution transport. Based on this JDA-POT adaptation model, it is supposed that negative transport effect will be removed and hence to improve the performance.
\section{Proposed unsupervised domain adaptation}
\label{sect_frm}
Our proposed UDA model is based on the current state of the art framework where X-vector extraction is used as a front-end model in the SLID system. The X-vector extraction is based on a deep neural network model, e.g., time delay neural network (TDNN) \cite{Snyder2018} or its extensions. Based on the X-vector, a feature projection (latent feature space) and classifier (label space) are designed. The model framework design is illustrated in Fig. \ref{figConv}. In this figure, the encoder block is for X-vector extraction, and is fixed in adaptation model training. The red dash-line box is the feature projection module where the X-vector is projected into a low-dimensional space for discriminative language feature extraction. In a conventional framework, it is composed of a linear discriminative analysis (LDA) and vector normalization. In our framework, a dense layer neural network model with vector length normalization (L-Norm) are used for this function. Based on the normalized feature, a discriminative classifier is designed as another dense layer neural network model with softmax activation (as showed in the blue dash-line box in Fig. \ref{figConv}). In the adaptation, only the feature projection and classifier with relatively small number of parameters are adapted while the parameters of X-vector extraction network are fixed.
\subsection{Basic domain adaptation problem}
\label{sect-cross}
\begin{figure}[tb]
\centering
\includegraphics[width=7cm, height=5cm]{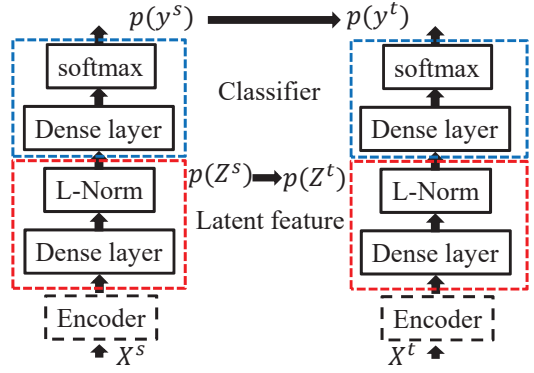}
\caption{Domain adaptation model framework for SLID based on X-vector extraction and classifier models}
\label{figConv}
\end{figure}

Given a source domain data set $D^s  = \left\{ {\left( {{\bf x}_i^s ,{\bf y}_i^s } \right)} \right\}_{i = 1,..,N}$, and a target domain data set $D^t  = \left\{ {\left( {{\bf x}_i^t ,{\bf y}_i^t } \right)} \right\}_{i = 1,..,M}$ (in real situations, the target label information is unknown). Due to domain changes (e.g. recording channels), there exists domain discrepancy, i.e., $p^s \left( {{\bf x},{\bf y}} \right) \ne p^t \left( {{\bf x},{\bf y}} \right)$. In most studies, the solution is to try to find a latent feature space with a transform ${\bf z} = \phi ({\bf x})$ by which the joint distribution could be approximated (or aligned) $p^s ({\bf z},{\bf y}) \approx p^t ({\bf z},{\bf y})$. Based on the Bayesian theory, the joint distribution of feature and label in the transformed space is formulated as:
{\small
\begin{equation}
p^u ({\bf z},{\bf y}) = p^u ({\bf y}|{\bf z})p^u ({\bf z});u = \{ s,t\}
\end{equation}
}
The joint distribution adaptation can be implemented as feature marginal distribution alignment and classifier conditional distribution alignment as:
{\small
\begin{equation}
\begin{array}{l}
 p^s ({\bf z}) \approx p^t ({\bf z}) \\
 p^s ({\bf y}|{\bf z}) \approx p^t ({\bf y}|{\bf z}) \\
 \end{array}
 \label{eq_joint}
\end{equation}
}
In Eq. (\ref{eq_joint}), the first formulation is for adaptation in covariance shift condition (as feature marginal distribution alignment). The second formulation is for adaptation in concept shift condition. In our study, a label marginal distribution alignment is applied instead, i.e., $p^s ({\bf y}) \approx p^t ({\bf y})$.
In our model framework, for training and test sharing the same neural network transforms, the adaptation is to minimize the distribution discrepancy of latent feature and classifier between training and test data set as illustrated in Fig. \ref{figConv}.
In order to measure the distribution discrepancy, optimal transport provides a powerful tool for measuring the distance between two distributions, i.e., optimal transport distance.
\subsection{Optimal transport distance}
For two marginal distributions from source $p^s$ and target $p^t$, the discrete OT distance (also known as Wasserstein distance) is defined as:
{\small
\begin{equation}
L_{{\rm OT}} (p^s ,p^t )\mathop  = \limits^\Delta  \mathop {\min }\limits_{\gamma  \in \prod (p^s ,p^t )} \sum\limits_{i,j} {L({\bf z}_i^s ,{\bf z}_j^t )\gamma ({\bf z}_i^s ,{\bf z}_j^t )}
\label{eq_ot}
\end{equation}
}
where ${\gamma  \in \prod {\left( {p^s ,p^t } \right)} }$ is the transport plan (or coupling) between the two distributions, and ${L({\bf z}_i^s ,{\bf z}_j^t )}$ is the transport cost between examples ${{\bf z}_i^s}$ and ${{\bf z}_j^t }$ that are sampled from marginal probability distributions $ p^s$ and $p^t$, respectively. The initial OT based adaptation is applied for finding a marginal latent feature space, and later it is modified for joint adaptation framework, i.e., both latent feature and classifier are adapted in image classification \cite{CourtyIEEE2017, CourtyNIPS2017, DamodaranECCV2018}. As in our task, labeling in target samples are only a subset of that of the source domain, therefore, in OT, it is better that only samples with labels shared by the testing data should be transported, i.e., partial optimal transport.
\subsubsection{Partial optimal transport}
\begin{figure}[tb]
\centering
\includegraphics[width=7cm, height=5cm]{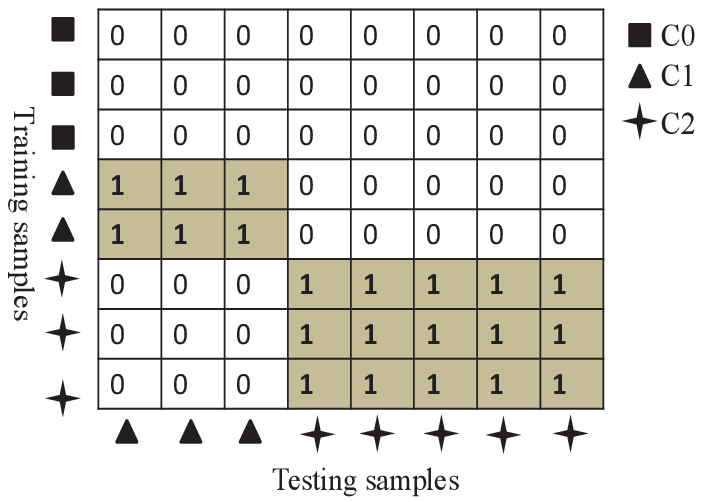}
\caption{Partial coupling in optimal transport (admissible transport pathes) between training (source) and testing (target) samples (refer to text for details).}
\label{figcoupling}
\end{figure}

In Eq. (\ref{eq_ot}), the OT distance is defined based on the assumption that the total probability mass of $p^s$ and $p^t$ are equal, and all probability mass from source $p^s$ is transported to target $p^t$. In real situations, when the label spaces of source and target domains are different, it is not suitable to transport all probability mass from source to the target. Therefore, the coupling function ${\gamma  \in \prod {\left( {p^s ,p^t } \right)} }$ should be constrained, i.e., POT by keeping only admissible couplings between probability mass induced by shared label classes in both source and target domains. The POT problem was discussed in \cite{Figalli2010}, and partial Wasserstein and Gromov-Wasserstein problems were addressed in \cite{Chapel2020}. Particularly, a joint POT has been recently proposed for open set domain adaptation problem \cite{JPOPT2021}. In all these studies, the probability masses of source and target domains are intrinsically different which should not be fully coupled in mass transportation. Based on this consideration, Eq. (\ref{eq_ot}) is changed to:
\begin{equation}
L_{{\rm POT}} (p^s ,p^t )\mathop  = \limits^\Delta  \mathop {\min }\limits_{\gamma  \in \prod ^{{\rm par}} (p^s ,p^t )} \sum\limits_{i,j} {L({\bf z}_i^s ,{\bf z}_j^t )\gamma ({\bf z}_i^s ,{\bf z}_j^t )}
\label{eq_pot}
\end{equation}
It is noted that in this POT based distance, the couplings ${\gamma  \in \prod ^{{\rm par}} (p^s ,p^t )}$ are only partial admissible between source and target domains. For convenience of understanding, we further illustrate the partial transport coupling between training and testing samples in Fig. \ref{figcoupling}.
In this figure, samples in vertical are from source domain with labeling set $\{ {\rm C0} ,{\rm C1} ,{\rm C2} \}$ (different symbols represent different classes as illustrated in Fig. \ref{figcoupling}), samples in horizontal are from target domain with labeling set $\{ {\rm C1} ,{\rm C2} \}$. The admissible couplings are marked with $1$ otherwise marked with $0$. Although this POT could easily remove the negative transfer effect by ignoring the non-admissible couplings, in real situations, the target labels are unknown, it is possible to set a threshold of the transport cost to obtain the admissible couplings as:
\begin{equation}
w_{i,j}  = \left\{ \begin{array}{l}
 1,L({\bf z}_i^s ,{\bf z}_j^t ) \le b \\
 0,L({\bf z}_i^s ,{\bf z}_j^t ) > b \\
 \end{array} \right.
 \end{equation}
where $b$ is the transport cost threshold, ${\bf z}_i^s$ and ${\bf z}_j^t$ are the two samples from source and target domains indexed by $i$ and $j$, respectively. In this equation, if the transport cost is smaller than $b$, the coupling between the two samples is allowed, otherwise, the coupling is discarded. However, this hard controlling of the couplings may have a risk in discarding the admissible or accepting non-admissible couplings. Therefore, a soft weighting on the coupling is applied which is defined as follows:
\begin{equation}
\tilde w_{i,j}  = \sigma \left( { - scale*\left( {L({\bf z}_i^s ,{\bf z}_j^t ) - b} \right)} \right),
\label{eq_softw}
\end{equation}
where $\sigma(.)$ is a sigmoid function, and $scale$ is a scaling parameter. Finally, the solution of the POT is changed to:
\begin{equation}
L_{{\rm POT}} (p^s ,p^t )\mathop  = \limits^\Delta  \mathop {\min }\limits_{\gamma  \in \prod (p^s ,p^t )} \sum\limits_{i,j} {L({\bf z}_i^s ,{\bf z}_j^t )\gamma ({\bf z}_i^s ,{\bf z}_j^t )\tilde w_{i,j} }
\label{eq_potfor}
\end{equation}
In this formulation, although the coupling ${\gamma  \in \prod (p^s ,p^t )}$ is the same as used in OT, a soft weighting defined in Eq. (\ref{eq_softw}) which can be regarded as the uncertainty of coupling is explicitly added to fulfil the function of POT.  In the followings, this POT based distance will be integrated in a neural network learning for cross-domain SLID.
 \subsection{Neural alignment model with OT and POT based loss}
 \label{subsect_NNOT}
In Fig. \ref{figConv}, the final output can be regarded as a composition of two functions defined as:
{\small
\begin{equation}
y\left( {\bf x} \right) = f\left( {{\bf x};\theta _g ,\theta _h } \right) = g \circ h\left( {\bf x} \right),
\label{eq_comp}
\end{equation}
}
where $h\left(  \cdot  \right)$ and $g\left(  \cdot  \right)$ are the feature extraction and classifier transforms with parameter sets ${\theta _h }$ and ${\theta _g }$, respectively. Correspondingly, the latent feature could be obtained as ${\bf z} = \phi \left( {\bf x} \right) = h\left( {\bf x} \right)$, and class-wise probability values could be obtained by the classifier module $g\left(  \cdot  \right)$. The adaptation process could be explicitly applied to the outputs of the two modules. With reference to the explanation in section \ref{sect-cross}, the joint cost function defined on latent feature and label for adaptation is defined as:
{\small
\begin{equation}
L_{{\rm ad}} ({\bf z}^s ,{\bf y}^s ;{\bf z}^t ,{\bf y}^t ) = \alpha L_{{\rm fea}} ({\bf z}^s ,{\bf z}^t ) + \beta L_{{\rm cls}} ({\bf y}^s ,{\bf y}^t ),
\label{eq_Ladapt}
\end{equation}
}
where $L_{{\rm fea}} ({\bf z}^s ,{\bf z}^t )$ is adaptation cost on latent feature distributions, and $L_{{\rm cls}} \left( {{\bf y}^s ,{\bf y}^t } \right)$ is adaptation cost on classifier or label distributions, $\alpha$ and $\beta$ are weighting coefficients for feature and classifier adaptation costs (simple Euclidian distance is used in this paper). Since label in target domain is usually unknown, an estimation ${\bf \hat y}^t  = f\left( {{\bf x}^t } \right) = g \circ h\left( {{\bf x}^t } \right)$ is used as target label, hence the adaptation cost defined in Eq. (\ref{eq_Ladapt}) is changed to $L_{{\rm ad}} ({\bf z}^s ,{\bf y}^s ;{\bf z}^t ,{\bf \hat y}^t )$. Using the cost defined in Eq. (\ref{eq_Ladapt}) as a transport cost function, for an unsupervised adaptation, the discrepancy measurement between source and target domains is defined as OT distance:
{\small
\begin{equation}
L_{{\rm OT}} (p^s ,p^t ) = \mathop {\min }\limits_{\gamma  \in \prod (p^s ,p^t )} \sum\limits_{i,j} {L_{{\rm ad}} ({\bf v}_i^s ,{\bf v}_i^t )\gamma ({\bf v}_i^s ,{\bf v}_j^t )},
\label{eq_OTAdpt}
\end{equation}
}
where ${\bf v}_i^s  \in \left\{ {({\bf z}_i^s ,{\bf y}_i^s )} \right\}$ and ${\bf v}_j^t  \in \left\{ {({\bf z}_j^t ,{\bf \hat y}_j^t )} \right\}$ are tuples of joint samples from source and target domains ($i$ and $j$ are sample indexes), respectively. Eq. (\ref{eq_OTAdpt}) is used to find the optimal transport plan matrix $\gamma$ by which the adaptation loss could be estimated. Correspondingly, with reference to POT definition in Eq. (\ref{eq_potfor}), the loss based on POT is defined as follows:
{\small
\begin{equation}
L_{{\rm POT}} (p^s ,p^t ) = \mathop {\min }\limits_{\gamma  \in \prod (p^s ,p^t )} \sum\limits_{i,j} {L_{{\rm ad}} ({\bf v}_i^s ,{\bf v}_i^t )\gamma ({\bf v}_i^s ,{\bf v}_j^t )\tilde w({\bf v}_i^s ,{\bf v}_j^t )},
\label{eq_POTAdpt}
\end{equation}
}
where $\tilde w_{i,j} (.)$ is estimated based on Eq. (\ref{eq_softw}) with cost defined in Eq. (\ref{eq_Ladapt}).
Besides the adaptation loss, the classification loss in source domain is defined as the multi-class cross entropy as:
{\small
\begin{equation}
L_{{\rm CE}}^s ({\bf y}_i^s ,{\bf \hat y}_i^s )\mathop  = \limits^\Delta   - \sum\limits_{j = 1}^{N_c } {y_{i,j}^s \log \hat y_{i,j}^s },
\end{equation}
}
where ${{\bf \hat y}_i^s }$ is the estimated label in source domain as ${\bf \hat y}_i^s  = f\left( {{\bf x}_i^s } \right) = g \circ h\left( {{\bf x}_i^s } \right)$ ($i$ as sample index, and $N_c$ is the number of class).
Therefore, the total loss including the adaptation loss and source domain classification loss is:
{\small
\begin{equation}
L_{\rm T}  = \mathop {\min }\limits_{\gamma ,\theta _g ,\theta _h } \left( {\sum\limits_i {L_{{\rm CE}}^s \left( {{\bf y}_i^s ,{\bf \hat y}_i^s } \right)}  + \lambda L_{{\rm OT}} \left( {p^s ,p^t } \right)} \right).
\label{eq_TLoss}
\end{equation}
}
By substituting $L_{{\rm OT}} (.)$ with $L_{{\rm POT}} (.)$ in Eq. (\ref{eq_TLoss}), we could obtain the total loss function based on POT. In optimization, $\gamma$, ${\theta _g }$, and ${\theta _h }$ in Eq. (\ref{eq_TLoss}) are involved. They could be alternatively estimated via an expectation-maximization (EM) like optimization framework with mini-batch sampling of source and target samples as introduced in \cite{CourtyNIPS2017,DamodaranECCV2018}.
\begin{figure*}[tb]
\centering
\includegraphics[width=13cm, height=5cm]{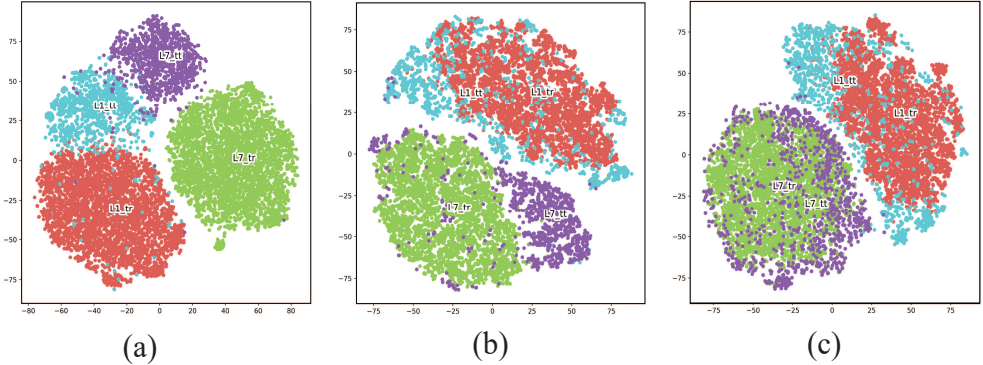}
\caption{Language cluster distributions: no adaptation (a), adaptation based on JDA-OT (b), and on JDA-POT (c). L1\_tr and L1\_tt: language 1 from training and test sets, respectively; L7\_tr and L7\_tt: language 7 from training and test sets, respectively.}
\label{fig_JBFrm}
\end{figure*}
\section{Experiments and results}
\label{sect_exp}
The data sets from Oriental Language Recognition (OLR) 2020 Challenge \cite{WD2019,WD2020,OLR2020} are used for examining the effectiveness of our proposed adaptation algorithm. The training set includes 110 k utterances ( more than 100 hours),  from 10 languages. And three test sets for two different tasks are provided, short utterance LID test in task 1, and cross channel LID test in task 2 with development and test sets.  Task 1 test set includes utterances from the same 10 languages of the training set (1.8 k utterances for each), but the utterance duration is short (1 s). Task 2 includes only 6 languages, also with 1.8 k utterances for each. But the utterances were recorded in wild environments which are quite different from those of for training data set. In task 2, there is one development set and one test set. In this paper, both are used as independent test sets for task 2. In order to measure the quality of the classification and adaptation models, two evaluation metrics are adopted by considering the missing and false alarm probabilities for target and nontarget language pairs, i.e., equal error rate (EER), and average performance cost (Cavg) as defined in \cite{WD2020}.

The X-vector extraction model is based on an extended TDNN architecture that was trained with the training set as well as augmented data \cite{OLR2020}. A little different from our previous study in \cite{ICASSPLu2021}, the input features for training the language embedding model are MFCCs extracted with 40 Mel bands. All other settings for X-vector extraction are the same as we used in \cite{ICASSPLu2021}. In model optimization, the Adam algorithm with an initial learning rate of $0.001$ was adopted \cite{Adam}, mini-batch size was 128. During model adaptation, the hyper-parameters were empirically set (with grid search method), transport cost threshold $b=1$, and scale parameter $scale=5.0$ (as defined in Eq. (\ref{eq_softw})), $\alpha=1.0$, $beta=0.001$ (defined in Eq. (\ref{eq_Ladapt})), and $\lambda=1.0$ (defined in Eq. (\ref{eq_TLoss})).

Based on the X-vector, the baseline performance without any adaptation methods are showed in Table \ref{tab1}.
\begin{table}[tbph]
\centering
\caption{Baseline performance (EER in (\%)).}
\begin{tabular}{|c||c|c|c|}
\hline
 Test sets&T1 &T2\_dev &T2\_test\\
\hline
\hline
EER &7.65&21.13&23.89\\
\hline
Cavg  &0.074&0.2128 &0.243\\
\hline
\end{tabular}
\label{tab1}
\end{table}

In Table \ref{tab1}, ``T1" is for short utterance LID test in task 1, ``T2\_dev" and ``T2\_test" are for cross-domain LID test in task 2 for develop and test sets, respectively. From these results, we can see that, the performance for cross-domain task was drastically degraded due to the domain mismatch problem. After the model was adapted, it is expected that the cluster distribution of each language is aligned to match between training and test sets. We first visually check the effect of unsupervised adaptation on language cluster distributions based on the t-Distributed Stochastic Neighbor Embedding (TSNE) \cite{Maaten2008}. The cluster distributions are showed in Fig. \ref{fig_JBFrm}. In this figure, samples of two languages from training and test sets are selected (labeled as L1\_tr, L1\_tt, L7\_tr, L7\_tt in the figure). From this figure, we can see that there is a large distribution shift between the training and test conditions for the baseline system without adaptation (Fig. \ref{fig_JBFrm}(a)). After adaptation with JDA-OT and JDA-POT ((Fig. \ref{fig_JBFrm}(b) and Fig. \ref{fig_JBFrm}(c)), the cluster of each different language is aligned to be overlapped. Moreover, it seems that the alignment based on JDA-POT is much better than that based on the JDA-OT. On the SLID task, the performance is showed in Table \ref{tab2}. Comparing the results in Tables \ref{tab1} and \ref{tab2}, we can see that with our proposed unsupervised adaptation learning, the performance for both task 1 and task 2 are improved. Particularly, with the newly proposed adaptation based on POT, the performance for cross-domain LID task 2 is improved with a large margin.
\begin{table}
\caption{Adaptation performance (EER in (\%))}
\centering
\begin{tabular}{|c|c|c|c|c|c|c|c|c|c|}
\hline
 & model & T1 & T2\_dev & T2\_test \\
\hline
\multirow{2}{*}{EER}  &JDA-OT   & 5.31 & 13.73 & 13.37   \\
&JDA-POT (proposed)  & \textbf{ 5.011} & \textbf{5.833} &  \textbf{6.602}  \\
\hline
\multirow{2}{*}{Cavg}  &JDA-OT  & 0.0524 & 0.1297 & 0.1333   \\
&JDA-POT (proposed)  &  \textbf{0.0503} & \textbf{0.0565} & \textbf{0.0661}  \\
\hline
\end{tabular}
\label{tab2}
\end{table}

\begin{figure}[tb]
\centering
\includegraphics[width=5cm, height=5cm]{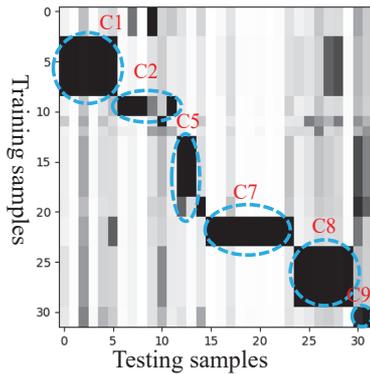}
\caption{Partial coupling weight in a mini-batch (32 training and 32 test samples) in adaptation.}
\label{figWeight}
\end{figure}

We further check the partial coupling weight used in Eq. (\ref{eq_POTAdpt}). After the adaptation model was trained, we selected 64 samples in a mini-batch, and sorted them with reference to their labels (in adaptation learning stage, target labels are unknown). The label space of training data is composed of 10 classes as \{C0-9\}, while the label space of test data only includes \{C1, C2, C5, C7, C8, C9\}. The partial coupling weights are showed in Fig. \ref{figWeight}. From this figure, we can see that partial couplings are clearly showed in the block structures between matched class samples (although there are sill a few couplings with mistakes).
\section{Conclusion}
\label{conclude}
In this paper, for cross-channel SLID tasks, we have previously proposed an JDA model based on OT for domain adaptation. Considering the label space of test set is only a subset of that of the training set, we further proposed an JDA model based on POT where only partial couplings were involved in distribution alignment. Our experimental analysis and results showed that the adaptation effectively improved the performance. Some problems are remained in this paper. One problem is that there are several hyper-parameters which were determined empirically, it is better that they are automatically learned during the optimization of the adaptation model. Moreover, as an adaptation method, there is a strong connection between our proposed one and other instance weighting based adaptation learning methods. In the future, we will further investigate these problems.
\bibliographystyle{IEEEtran}

\begin{thebibliography}{9}
\bibitem{Snyder2018}
D. Snyder, D. Garcia-Romero, G. Sell, D. Povey, and S. Khudanpur, ``X-vectors: Robust dnn embeddings for speaker recognition," in \emph{Proc. of ICASSP}, pp. 5329-5333, 2018.

\bibitem{RichardsonIEEE}
R. Richardson, D. Reynolds, N. Dehak, ``Deep Neural Network Approaches to Speaker and Language Recognition," \emph{IEEE Signal Processing Letters}, 22 (10), 1671-1675, 2015.

\bibitem{Moreno2016}
I. Lopez-Moreno, J. Gonzalez-Dominguez, D. Martinez, O. Plchot, J. Gonzalez-Rodriguez, P. J. Moreno, ``On the use of deep feedforward neural networks for automatic language identification," \emph{Computer Speech \& Language}, vol.40, pp.46-59, 2016.

\bibitem{Moreno2014}
I. Lopez-Moreno, J. Gonzalez-Dominguez, O. Plchot, D. Martinez, J. Gonzalez-Rodriguez and P. Moreno, "Automatic language identification using deep neural networks," in \emph{Proc. of ICASSP}, pp. 5337-5341, 2014.

\bibitem{Diez2015}
A. Lozano-Diez, R. Zazo Candil, J. G. Dominguez, D. T. Toledano and J. G. Rodriguez, ``An end-to-end approach to language identification in short utterances using convolutional neural networks," in \emph{Proc. of INTERSPEECH}, pp. 403-407, 2015.

\bibitem{Fernando2017}
 S. Fernando, V. Sethu, E. Ambikairajah and J. Epps, ``Bidirectional Modelling for Short Duration Language Identification," in \emph{Proc. of INTERSPEECH}, pp. 2809-2813, 2017.

\bibitem{Geng2016}
W. Geng, W. Wang, Y. Zhao, X. Cai and B. Xu, ``End-to-End Language Identification Using Attention-Based Recurrent Neural Networks," in \emph{Proc. of INTERSPEECH}, pp. 2944-2948, 2016.

\bibitem{LeeICASSP2019}
K. Lee, Q. Wang, and T. Koshinaka, ``The coral+ algorithm for unsupervised domain adaptation of plda," in  \emph{Proc. of ICASSP}, pp. 5821-5825, 2019.

\bibitem{Bousquet2019}
P. Bousquet, M. Rouvier, ``On robustness of unsupervised domain adaptation for speaker recognition," in \emph{Proc. of INTERSPEECH}, pp. 2958-2962, 2019.

\bibitem{MMD}
B. Scholkopf, J. Platt, T. Thomas, ``A Kernel Method for the Two-Sample-Problem," in \emph{Proceedings of the Conference of Advances in Neural Information Processing Systems}, vol. 19, pp.513-520, 2006.

\bibitem{Ganin2015}
Y. Ganin, E. Ustinova, H. Ajakan, P. Germain, H. Larochelle, F. Laviolette, M. Marchand, and V. Lempitsky, ``Domain-adversarial training of neural networks," \emph{Journal of Machine Learning Research},  vol. 17, no. 1, pp. 2096-2030, 2016.

\bibitem{CourtyIEEE2017}
N. Courty, R. Flamary, D. Tuia, A. Rakotomamonjy, ``Optimal transport for domain adaptation," \emph{IEEE Transactions on Pattern Analysis and Machine Intelligence}, 39 (9), pp. 1853-1865, 2017.

\bibitem{Peyre2018}
G. Peyre, M. Cuturi, ``Computational Optimal Transport," ArXiv:1803.00567, 2018.

\bibitem{CourtyNIPS2017}
N. Courty, R.  Flamary, A. Habrard, A. Rakotomamonjy, ``Joint distribution optimal transportation for domain adaptation," In \emph{Proc. of NIPS}, pp. 3730-3739, 2017.

\bibitem{DamodaranECCV2018}
B. Damodaran, B. Kellenberger, R. Flamary, D. Tuia, and N. Courty, ``DeepJDOT: Deep joint distribution optimal transport for unsupervised domain adaptation," In \emph{Proceedings of the European Conference on Computer Vision (ECCV)}, pp. 447-463, 2018.

\bibitem{LinNIPS2021}
H. Lin, H. Tseng, X. Lu, Y. Tsao, ``Unsupervised Noise Adaptive Speech Enhancement by Discriminator-Constrained Optimal Transport," \emph{Advances in Neural Information Processing Systems (NeurIPS)}, vol. 34, 2021.

\bibitem{ICASSPLu2021}
X. Lu, P. Shen, Y. Tsao, H. Kawai, ``Unsupervised Neural Adaptation Model Based on Optimal Transport for Spoken Language Identification,"  in \emph{Proc. of ICASSP}, pp. 7213-7217, 2021.

\bibitem{Figalli2010}
A. Figalli, ``The Optimal Partial Transport Problem," \emph{Arch Rational Mech Anal} 195, pp. 533-560, 2010.

\bibitem{Chapel2020}
L. Chapel, M. Alaya, and G. Gasso, ``Partial optimal transport with applications on positive unlabeled learning," \emph{In Advances in Neural Information Processing Systems (NeurIPS)}, vol. 33, pp. 2903-2913, 2020.

\bibitem{JPOPT2021}
R. Xu, P. Liu, Y. Zhang, F. Cai, J. Wang, S. Liang, H. Ying, J. Yin,  ``Joint partial optimal transport for open set domain adaptation," in \emph{Proceedings of the Twenty-Ninth International Joint Conference on Artificial Intelligence}, no. 352, pp. 2540-2546, 2021.

\bibitem{WD2019}
Z. Tang, D. Wang, L. Song, ``AP19-OLR Challenge: Three Tasks and Their Baselines," in \emph{Proc. of APSIPA ASC}, pp. 1917-1921, 2019.

\bibitem{WD2020}
Z. Li, M. Zhao, Q. Hong, L. Li, Z. Tang, D. Wang, L. Song, C. Yang, ``AP20-OLR Challenge: Three Tasks and Their Baselines," in \emph{Proc. of APSIPA ASC}, pp. 550-555, 2020.

\bibitem{OLR2020}
http://cslt.riit.tsinghua.edu.cn/mediawiki/index.php/OLR\_Challenge\_2020

\bibitem{Maaten2008}
L. Maaten, G. Hinton, ``Visualizing Data Using t-SNE," \emph{Journal of Machine Learning Research}, 9 (86), pp. 2579-2605, 2008.

\bibitem{Adam}
Diederik P. Kingma, Jimmy Ba, ``Adam: A Method for Stochastic Optimization," \emph{the 3rd International Conference on Learning Representations (ICLR}), 2014.

\end{thebibliography}

\end{document}